\documentclass{acmsiggraph}                     % final
\usepackage{booktabs} % For formal tables
\usepackage{microtype}
\usepackage{amsmath}
\usepackage{amssymb}
\usepackage{units}
\DeclareMathAlphabet{\mathcal}{OMS}{cmsy}{m}{n}  % reset mathcal font
\usepackage{xfrac}
\usepackage[normalem]{ulem}
\usepackage{graphicx}  		% already in smrdefault
\usepackage{float}
\usepackage{url}
\usepackage{xspace}
\usepackage{color}
\usepackage[normalem]{ulem}  % for strike-through (\sout)  % used by Tim's macros
\usepackage{enumitem}  		% to support [resume]
\usepackage{tikz}			% support for drawing shapes
\usepackage{calc}			% support adding lengths together
\usepackage{collcell}
\usepackage{tabularx}

\usepackage{multirow}
\usepackage{amsmath}
\usepackage{amssymb}
\usepackage{amsfonts}
\usepackage{eucal}
\usepackage[latin1]{inputenc}
\usepackage[normalem]{ulem}
\usepackage{listings}
\usepackage{mathrsfs}
\usepackage{subcaption}
\usepackage{cancel}
\usepackage[normalem]{ulem}
\usepackage{cancel}
\usepackage{tcolorbox}
\usepackage{verbatimbox}

\usepackage{graphicx}
\usepackage{color}
\usepackage{wrapfig}
\usepackage{ifthen}
\usepackage[]{algorithm2e}
\usepackage{algorithmicx}
\usepackage{algpseudocode}

\DeclareMathOperator*{\argmin}{arg\,min}

\title{Blue-Noise Dithered QMC Hierarchical Russian Roulette}
\author{Jacopo Pantaleoni\thanks{e-mail: jpantaleoni@nvidia.com}\\NVIDIA}


\begin{document}
	
	\newcommand{\picresdir}{final}

	\teaser{
		\includegraphics[width=163.0mm]{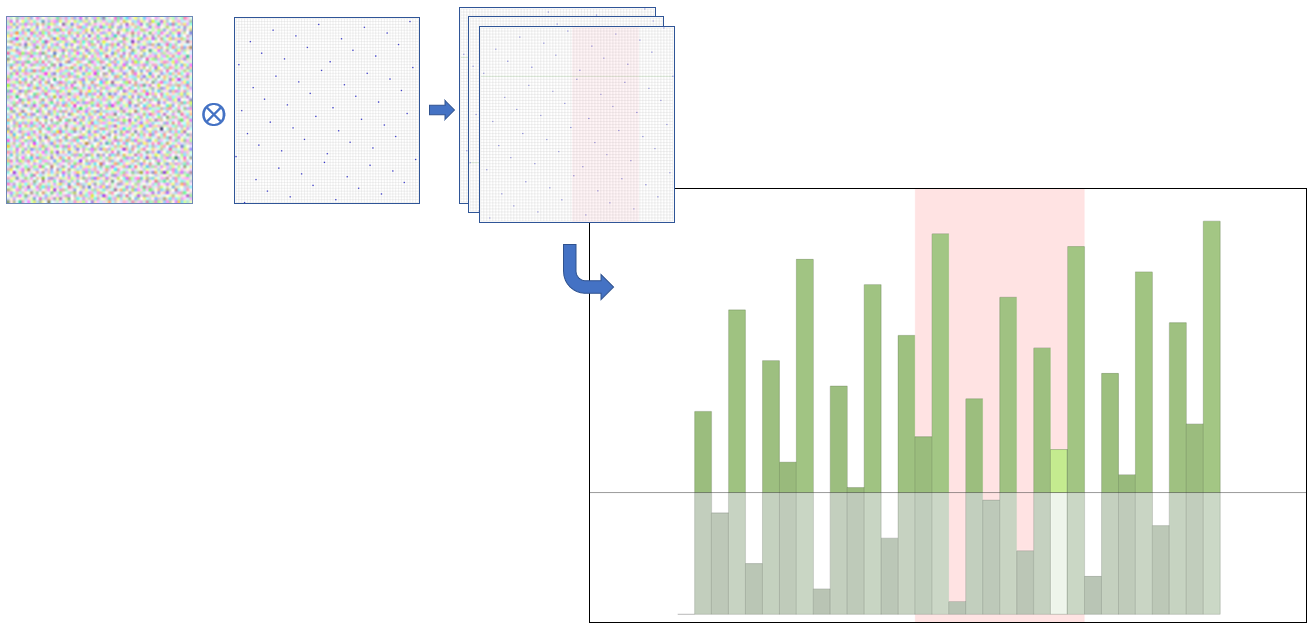}
		\caption{
			This work replaces the pseudo-random sequences used in the original hierarchical Russian roulette many-light algorithm with blue-noise dithered QMC sequences, making use of a novel algorithm to compute the minimum of a shifted Halton sequence over arbitrary intervals.
A tile of blue-noise distributed values (top row, left) is associated with the 2d pixel distribution of the eye vertices; each pixel in the tile is used to shift a common base-2 Halton sequence (top row, middle), producing a multitude of shifted sequences, one for each eye vertex $e_j$ (top-row, right). The green horizontal lines in the shifted sequences represent the shift values.
			The figure on the bottom shows a representation of the key operation needed to perform hierarchical Russian roulette, finding the minimum over a range of indices, highlighted in pink: notice that since the shift is performed modulo 1, if the shift is $r$, in the unshifted sequence the minimum value has to be found in the vertical range $[1-r, 1)$.
%			This paper addresses the efficient computation of the minimum value of a Cranley-Patterson rotated radical inverse sequence across an arbitrary range of integers $[a,b)$ - useful to perform hierarchical Russian Roulette.
%			In this picture, each bar represents the value of the radical inverse function at a given integer index (alternatively, their top-left corners can be thought of as representing a Hammersley QMC point-set), while the query range is highlighted in pink; moreover, since the shift in the Cranley-Patterson rotation is performed modulo one, a shift by $r$ induces a threshold $1 - r$ in the unshifted graph, above which samples assume a lower shifted value, indicated by the horizontal line and more saturated bar colors. The point with minimum value is highlighted in bright green.
		}
	}
	
	%% The ``\maketitle'' command must be the first command after the
	%% ``\begin{document}'' command. It prepares and prints the title block.
	
	\maketitle
	
	%% Abstract section.
	
	%% ACM Computing Review (CR) categories. 
	%% See <http://www.acm.org/class/1998/> for details.
	%% The ``\CRcat'' command takes four arguments.
	
	%\begin{CRcatlist}
	%	\CRcat{I.3.2}{Graphics Systems C.2.1, C.2.4, C.3)}{Stand-alone systems};
	%	\CRcat{I.3.7}{Three-Dimensional Graphics and Realism}{Color,shading,shadowing, and texture}{Raytracing};
	%\end{CRcatlist}

	%% The ``\keywordlist'' command prints out the keywords.
	\keywordlist
	
	%% The ``\copyrightspace'' command must be the first command after the 
	%% start of the first section of the body of your paper. It ensures the
	%% copyright space is left at the bottom of the first column on the first
	%% page of your paper.
	
	\copyrightspace

\abstract

In order to efficiently sample specular-diffuse-glossy and glossy-diffuse-glossy transport phenomena,  \cite{Tokuyoshi:2019:HRR} introduced \emph{hierarchical Russian roulette}, a smart algorithm that allows to compute the minimum of the random numbers associated to the leaves of a tree at each internal node.
The algorithm is used to efficiently cull the connections between the product set of eye and light vertices belonging to large caches of eye and light subpaths produced through bidirectional path tracing \cite{Veach:PHD}.

The original version of the algorithm is entirely based on the generation of semi-stratified pseudo-random numbers.
Our paper proposes a novel variant based on deterministic blue-noise dithered Quasi Monte Carlo samples \cite{Georgiev:2016}.

\section{Introduction}

Throughout this work, we assume that we have two large sets of eye vertices, $\{e_j : j \in [0, N_e) \}$ and light vertices $\{l_i : i \in [0, N_l) \}$.
We further assume that the set of light vertices is sorted by a hierarchy, so that each leaf is given an index $i$ in $[0,N_l)$, and each internal node $n$ of the hierarchy can be represented by a sub-range $[a_n,b_n) \subseteq [0,N_l)$.

Tokuyoshi and Harada \shortcite{Tokuyoshi:2019:HRR} showed that if one associates a randomly sized culling shape to each pair of eye and light vertices $(e_j, l_i)$, where the size of the culling shape is inversely proportional to a pseudo-random number $\xi_{ij}$, it is possible to use the hierarchy over the lights to efficiently cull all lights in sub-linear time for each eye vertex, performing so called \emph{hierarchical Russian roulette}.
The key operation required to perform hierarchical Russian roulette is a top-down tree traversal where the minimum of all pseudo-random numbers within each node's range is efficiently computed.

In this work, we replace the pseudo-random sequences with deterministic QMC sequences.
Specifically, we assign a unique Cranley-Patterson rotated Halton sequence to each eye vertex $j$:

\begin{equation}
\phi_j(i) = (\Phi_2(i) + r_j) \mod 1
\end{equation}

where $\Phi_2$ is the base-2 radical inverse (also known as van der Corput sequence, see \cite{Niederreiter:1992:RNG:130653}) and $r_j$ is the Cranley-Patterson rotation, or \emph{shift}, assigned to the given eye vertex. In practice, if the eye vertex indices are sorted by pixel and path depth, the generation of the shifts can be easily made such that its image-plane projection has a desirable blue-noise distribution \cite{Georgiev:2016}.
Notice that the sequence index $i$ here corresponds to a light vertex index in the range $[0,N_l)$: hence, if the light vertices are spatially sorted, as for example done in Matrix Bidirectional Path Tracing \shortcite{Chaitanya:2018:MBPT}, the low-discrepancy and stratification properties of the sequence will be inherited by the distribution of the final samples.
%The process of blue-noise dithering is shown in Figure~\ref{BlueNoiseDithering}.

Now, given the above setting, the key operation needed to perform hierarchical Russian roulette consists in finding, for each internal node, the minimum value that the sequence assumes inside its range $[a,b)$:
\begin{equation}
v(a,b) = \min_{i \in [a,b) } { \phi_j(i) }
\end{equation}
While for the final application the use of this novel sampling scheme has interesting aspects of its own, the focus of this work will be the solution of this central problem.

%\begin{figure}
%	\includegraphics[width=82.0mm]{blue_noise_dithering}
%	\caption{A tile of blue-noise distributed values (left) is associated with the 2d pixel distribution of the eye vertices; each pixel in the tile is used to shift a common base-2 Halton sequence (middle), producing a multitude of shifted sequences, one for each eye vertex $e_j$. The green horizontal lines in the shifted sequences represent the shift values.}
%	\label{BlueNoiseDithering}
%\end{figure}

\section{Computing minima of the shifted Halton sequence over arbitrary intervals}

We first recap the definition of the radical inverse:

\begin{eqnarray}
\Phi_b(i) : \mathbb{N} &\rightarrow& \mathbb{Q} \cap [0,1) \nonumber \\
i = \sum_k a_k(i) b^k &\mapsto& \sum_k a_k(i) b^{-k-1}
\end{eqnarray}
In practice, for any index $i$, the radical inverse is obtained by first reversing the binary expansion of $i$, and then using that as the binary expansion of the base-2 digits after the comma of the result.

We will also change the problem setting by trying to find the index $k \in [a,b)$ that minimizes $\phi_j$:

\begin{equation}
k = \argmin_{i \in [a,b) } { \phi_j(i) }
\end{equation}
In the absence of a shift, $\phi_j = \Phi_2$, and since the most important digits of $i$ become the least important in the expansion of $\Phi_2(i)$, finding $k$ would mean finding the number in $[a,b)$ with the most rightmost zeros.
This can be done with an $O(B)$ algorithm in the number of bits $B$, which we report in Algorithm~1.

The presence of the shift makes the problem slightly more complex. Pseudo-code for the solution we are about to detail is given in Algorithm~2.
In the following, to simplify notation we'll drop the explicit dependence on the eye vertex index, and simply use $r$ instead of $r_j$.
Now, we can see that for a given sequence index $i$ there are essentially three cases:
\begin{eqnarray}
\Phi_2(i) &=& 1 - r \Rightarrow \phi_j(i) = 0 \\
\Phi_2(i) &>& 1 - r \Rightarrow \phi_j(i) \in (0, r) \\
\Phi_2(i) &<& 1 - r \Rightarrow \phi_j(i) \in (r,1)
\end{eqnarray}

The first case can only be obtained if $k_r := \Phi_2^{\leftarrow}(1 - r) \in [a,b)$. If that is the case, $k = k_r$ and the minimum value assumed by $\phi_j$ inside the node range is zero.

\RestyleAlgo{boxruled}
\begin{algorithm}%[frame=lines,label=inversion,caption={overall algorithm},captionpos=b]]
\SetStartEndCondition{ }{}{}%
\SetKwProg{Fn}{def}{\string:}{}
\SetKwFunction{Range}{range}%%
\SetKw{KwTo}{in}\SetKwFor{For}{for}{\string:}{}%
\SetKwIF{If}{ElseIf}{Else}{if}{:}{elif}{else:}{}%
\SetKwFor{While}{while}{:}{fintq}%
\AlgoDontDisplayBlockMarkers\SetAlgoNoEnd\SetAlgoNoLine%

	\SetKwFunction{fls}{fls}
	\SetKwFunction{ffs}{ffs}
	\SetKwFunction{clear}{clear}
	\SetKwFunction{set}{set}
	\SetKwFunction{isSet}{is\_set}
	{
		\If {$a == 0$ \textbf{or} a == b-1} {
			\textbf{return} a\;
		}

		\vspace{2mm}
		// find the most significant bit set of $b$ \\
		$b_h$ = \fls{b}\;

		\vspace{2mm}
		// find the least significant bit set of $a$\\
		$b_l$ = \ffs{a}\;

		\vspace{2mm}
		// sweep right-to-left, each time clearing \\
		// the current bit and setting the next\\
		$k = m = a$\;
		\For {$i = b_l$ \KwTo $b_h - 1$}{
			$m$  = \clear{$m$, $i$}; // clear the i-th bit \\

			\vspace{2mm}
			$m$  = \set{$m$, $i+1$}; // set bit i+1 \\

			\vspace{2mm}
			\If {$m \in [a,b)$} {
				$k = m$\;
			}
		}
		\textbf{return} $k$\;
	}
	\caption{find the number in $[a,b)$ with most rightmost zeros}
\end{algorithm}

If $k_r \notin [a,b)$, the next option is to look for an index $i$ such that its radical inverse $\Phi_2(i)$ is the smallest number larger than 1 - r.
That is to say, we'd have to look for solutions of:
\begin{equation}
k = \argmin_i \big\{ \Phi_2(i) \big\} \, \, : \, \, \ i \in [a,b) \, \, \cap \, \, \{ \Phi_2(i) > 1 - r \}
%k = \argmin_{i \in [a,b) \, : \, \Phi_2(i) > 1 - r } \left( { \Phi_2(i) \, : \, \Phi_2(i) > 1 - r } \right)
\end{equation}
In order to have a value of $\Phi_2(k)$ that is larger than $1 - r$, but at the same time as small as possible, we'll start looking at the bit structure of $a$, $b$ and $k_r$.
We'll indicate the position of the last (i.e. highest) bit set of $b$ with $b_h = fls(b)$, and the first (i.e. lowest) bit set of $k_r$ with $k_p = ffs(k_r)$.
If $k_p$ is part of a contiguous range of bits set in $k_r$, we'll indicate the entire range with $\{ k_p,...,k_q\}$, with $k_p < ... < k_q$.
(e.g. $b = 0{\bf 1}0010$, we'll have $b_h = 4$ and if $k_r = 100{\bf 1}{\bf 1}0$, we'll have $p = 1$, and $q = 2$).
Moreover, we'll indicate the last bit set of $a$ with $a_h = fls(a)$.We're now ready to proceed with the algorithm descriptions.

\subsection{Left-to-right sweep}
We'll start by considering a mask $m$ obtained by clearing all the leading bits of $k_r$ higher than $b_h$ (e.g. with $k_r$ and $b$ as in the example above, we'd consider $m = 000110$).
The main idea is that we'll try to craft $k$ by flipping only the highest possible zero bit of $m$ to the left of $k_q$ that leads to a value contained in $[a,b)$. We do this, because the highest bits of $m$ are the least important in its image $\Phi_2(m)$, due to the bit reversal in the radical inverse, and by modifying them we are trying to obtain a value that is just slightly larger than $1 - r$.

Hence, we'll proceed with a single left-to-right sweep through its bits, $b_i$, starting from $b_h$ and ending at $k_p$, performing the following sequence of operations:
\begin{itemize}
\item if the bit $b_i$ is already set in $m$, we clear it and increment $i$, continuing to the next bit;
\item else, we set the corresponding bit in $m$, obtaining:
	\begin{equation}
	m_i = m \, | \, 2^{b_i};
	\end{equation}
	by construction, $\Phi_2(m_i) > 1 - r$; hence, if $m_i$ is contained in the range $[a,b)$, and if $a_h \leq i < b_h$ we can conclude that $k = m_i$;
\end{itemize}

If during the loop we moved beyond $a_h$, it might still be possible to find a valid index satisfying equation (6) by altering a single bit between $a_h$ and $k_q$; however,
we need to make sure we keep and merge into $m$ all the set bits of $a$ we encounter in our left-to-right walk, as otherwise we'd produce an index smaller than $a$, and hence outside the target range; in fact, occasionally, if the $i$-digit prefix of $m$ (and hence $k_r$) is smaller than the corresponding prefix of $a$, even after merging all the bits of $a$ into $m$, $m_i$ might be too small to fall into $[a,b)$: in these cases, we set an extra $1$ bit at index $i+1$;
it is important to note, however, that in this case $m_i$ is no longer necessarily the smallest such value: we are only guaranteed that the bit $b_i$ is the highest possible bit in $k_r$ (i.e. the lowest possible bit in the binary expansion of $1 - r$) at which a flip from zero to one produces a value inside $[a,b)$ satisfying equation (6), or - in other words - that the minimum value $k$ exists and must have bit $b_i$ set.

\subsubsection{Solution refinement}

In order to find $k$ given $m_i$ we'll have to make several observations. The first is that the \emph{prefix} of $k$, starting at bit $0$ and ending at bit $i$, is now fixed as $k^* = \{ {\bf 1} k_{i-1} ... k_0 \}$, where $k_j$ is the $j$-th bit of $k_r$.
In the following, we'll indicate the (i+1) digit prefix of the binary expansion of a number $n$ with $n[0,i]$.
We'll also call $I$ the subset of $[a,b)$ with prefix $k^*$, i.e. $I = \{ j \in [a,b) \, \, : \, \, j[0,i] = k^*\}$, and indicate with $I_0$ the subset of $I$ which has bit (i+1) set to 0, and $I_1$ the subset of $I$ with bit (i+1) set to 1, and similarly call $I_{00}$ the subset of $I_0$ with bit (i+2) set to 0, and so on. We'll then have the following relations:
\begin{eqnarray}
	\Phi_2(k_1) < \Phi_2(k_2) \, \, \, &\forall& \, \, k_1 \in I_0, \, \, \, k_2 \in I_1 \nonumber \\
	\Phi_2(k_1) < \Phi_2(k_2) \, \, \, &\forall& \, \, k_1 \in I_{00}, \, \, \, k_2 \in I_{01} \nonumber \\
	\Phi_2(k_1) < \Phi_2(k_2) \, \, \, &\forall& \, \, k_1 \in I_{000}, \, \, \, k_2 \in I_{001}
\end{eqnarray}
In other words, we can establish an ordering relation among the sets $I_{j_1 j_2 ... j_n}$:

\begin{equation}
I_{s_1 s_2 ... s_n} < I_{t_1 t_2 ... t_n} \Leftrightarrow {s_1 s_2 ... s_n} < {t_1 t_2 ... t_n}.
\end{equation}

Now, if the prefix $b[0,i]$ is larger than $k^*$, we can form $k$ as a number that has 1 in position $b_h$ and zeros for each index $j : b_i < j < b_h$, i.e. $k = b_h 0 ... 0 k^*$ and conclude our search, as $k$ belongs to the \emph{smallest} set $I_{0 ... 01}$. If that is not the case, we'll have to proceed by induction, and start by looking for a solution in $I_0$. If that exists, we'll proceed to $I_{00}$. If not, we'll look for one in $I_{10}$. And so on.

By construction, the number $m_i$ we obtained might already belong to a given set $I_{0 ... 01 ...}$, with a given number $n_0$ of zeros before the first bit set is encountered. In that case, we'll directly skip those $n_0$ zero bits, and simply assume, without lack of generality, they are part of $k^*$. With this assumption, we know that the next bit in position $i+1$ must be 1 in $m_i$. Let's assume we have a whole sequence of 1's, $m_i[i+1, i+j] = 1...1$. In this case, our only chance to find a number in $I_0$ is to flip all those 1's to 0's, and change the next bit to 1, which can be achieved by means of a single addition of $2^{i+1}$. This is because this procedure will produce the smallest number bigger than our current value of $m_i$ having the bit in position $i+1$ set to $0$.
If the value thus obtained belongs to $[a,b)$, we are done, as we effectively found a number in $I_{s_1...s_{j+1}}$, with $s_1 = ... = s_j = 0$ and $s_{j+1} = 1$.
Now we can repeatedly apply the same algorithm to the next sequence of 1's.
Pseudo-code for this procedure is given in Algorithm 3.

\subsection{Right-to-left sweep}

If we conclude the left-to-right loop without finding $k$, it means it is only possible to satisfy equation (6) by flipping bits of $m$ to the right of $k_p$.
For example, this would happen with $a = 000101$, $b = 001000$, and $k_r = 010110$: by necessity, $k$ would have to have the form *****${\bf 1}$.
In this case, we'll proceed with the same right-to-left sweep used in Algorithm~1, except we modify its bounds and we further track the case $\Phi_2(k) > 1 - r$, which is only fulfilled if the last bit set to 1 preceeds $k_p$. If no such $k$ is found, it simply means that there is no index $i$ in $[a,b)$ such that $\Phi_2(i) > 1 - r$, and the algorithm proceeds till the maximum word size $B$ is reached, allowing it to automatically catch corner cases like $a = 0$.

\begin{figure}
	\includegraphics[width=82.0mm]{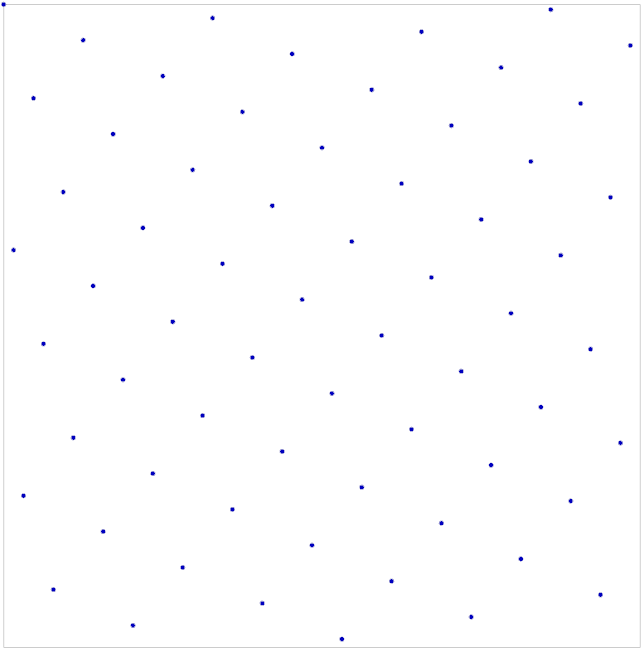}
	\caption{The golden-ratio rank-1 lattice can be used to further shift the base sequence to obtain multiple samples for each eye vertex.}
	\label{GoldenRatio}
\end{figure}

\section{Adaptive sampling}

The sequence definition we proposed uses the light hierarchy leaf index to directly address into a sequence, resulting in a single sample per light. With this setup, the resulting samples are stratified across all lights, but it is not directly possible to extend the process and its stratification to multiple samples per light.

In order to achieve this, we here propose a simple variant.
The idea is to use another sequence $\psi(n)$ to rotate the first, while stratifying across $n$. That is to say, we redefine $\phi_j$ as:

\begin{equation}
\phi_j(i,n) = \phi_j(i) +  \psi(n) \mod \, 1
\end{equation}

Note that we're free to choose any sequence for $\psi$, possibly even depending on $j$. From the purpose of the application of our algorithm, for a fixed $n$, nothing is changed,
as the new term can be absorbed into the shift:

\begin{equation}
\phi_j(i,n) = \Phi_2(i) +  (r_j + \psi(n)) \mod \, 1
\end{equation}

In practice, we suggest using a simple low dimensional sequence with desirable blue-noise and low-discrepancy distribution, such as the rank-1 lattice:

\begin{equation}
\psi_(n) = s + n \cdot \alpha \mod \, 1
\end{equation}
where $s$ is an optional shift (potentially set to zero) and $\alpha$ is any irrational number.
The choice of $\alpha$ that achieves the lowest possible discrepancy is $\alpha = 1 / \phi$, where $\phi$ is the well known golden ratio $(\sqrt{5} + 1) / {2} \approxeq 1.61803398875$ (see Figure~\ref{GoldenRatio}).

Another option is to get rid of the explicit shifts $r_j$ and use a single higher-dimensional sequence $\psi(j,n)$ that is stratified and has blue-noise properties both in space (i.e. across pixels) and in time:
\begin{equation}
\phi_j(i,n) = \Phi_2(i) +  \psi(j,n) \mod \, 1
\end{equation}
For small $n$, this could for example be achieved with a 3d tile of precomputed shifts. Alternatively, $\psi$ could be crafted with the techniques introduced by Heitz and Belcour \shortcite{Heitz:2019:MCBN}.

\RestyleAlgo{boxruled}
\begin{algorithm}%[frame=lines,label=inversion,caption={overall algorithm},captionpos=b]]
\SetStartEndCondition{ }{}{}%
\SetKwProg{Fn}{def}{\string:}{}
\SetKwFunction{Range}{range}%%
\SetKw{KwTo}{in}\SetKwFor{For}{for}{\string:}{}%
\SetKwIF{If}{ElseIf}{Else}{if}{:}{elif}{else:}{}%
\SetKwFor{While}{while}{:}{fintq}%
\AlgoDontDisplayBlockMarkers\SetAlgoNoEnd\SetAlgoNoLine%

	\SetKwFunction{fls}{fls}
	\SetKwFunction{ffs}{ffs}
	\SetKwFunction{clear}{clear}
	\SetKwFunction{set}{set}
	\SetKwFunction{isSet}{is\_set}
	\SetKwFunction{refine}{refine}
	{
		// first case \\
		\If{$k_r \in [a,b)$} {
			\textbf{return} $k_r$\;
		 }		

		%\If{$a == b-1$} {
		%	\textbf{return} $a$\;
		%}		

		\vspace{2mm}
		// find the most significant bit set of $b$ \\
		$b_h$ = \fls{b}\;

		\vspace{2mm}
		// find the most significant bit set of $a$ \\
		$a_h$ = \fls{a}\;

		\vspace{2mm}
		// find least significant range of bits set of $k_r$ \\
		$k_p$ = \ffs{$k_r$}\; 
		$k_q$ = $k_p$\; 
		\While{\isSet{$k_r$, $k_q+1$}}{
			$k_q = k_q+1$\;
		}

		\vspace{2mm}
		// start building $m$ \\
		m = $k_r$\;

		\vspace{2mm}
		// clear all its bits more important than $b_h$ \\
		\For {$i > b_h$}{
			$m$  = \clear{$m$, $i$}\;
		}

		\vspace{2mm}
		// start our left-to-right sweep \\	
		$i$ = $b_h$\;
		\While{$i \geq k_p$}{
			\If{\isSet{ $m$, $i$\,}} {
				$m$  = \clear{$m$, $i$};  // clear the i-th bit \\
			 }
			\Else {
				$m_i$ = \set{$m$, $i$}; // set the i-th bit \\

				\vspace{2mm}
				\If{$m_i < a$}
				{
					 // try to make $m_i$ a little bigger \\
					$m_i$ = \set{$m_i$, $i+1$}\;
				}
				\vspace{2mm}
				$m_i$ = \refine{ $m_i, i+1, [a,b)$\,}\;

				\vspace{2mm}
				\If{$m_i \in [a,b)$}{
					\textbf{return} $m_i$\;
				}

%				\If{$m_i \in [a,b)$}{
%					\textbf{return} \refine{ $m_i, i+1, [a,b)$\,}\;
%				}
%				\If{$m_i < a$}
%				{
%					 // try to set the next bit to make this a little bigger... \\
%					$k$ = \set($m_i$, $i+1$)\;
%					$k$ = \refine{ $a, b, k, i+1$\,}\;
%					\If{$k \in [a,b)$}{
%						\textbf{return} k\;
%					}
%				}
			}

			\vspace{2mm}
			\If{\isSet{ $a$, $i$ }} {
				$m$  = \set{ $m$, $i$\,};  // keep the i-th bit from a \\
			 }
			$i$ = $i$ - 1\;
		}

		\vspace{2mm}
		// sweep right-to-left, each time clearing \\
		// the current bit and setting the next\\
		$m = a$; // start from m = a \\
		$k = b$; // mark k as invalid \\
		\For {$i = 0$ \KwTo $B-1$}{
			// if $i == k_p$ and we found a valid $k \in [a,b)$ \\
			// we can stop here: by looking further we will not \\
			// find any other with $\Phi_2(k) > 1 - r$ \\
			\If {$k \in [a,b)$ \textbf{and} $i == k_p$} {
				\textbf{return} $k$\;
			}

			\vspace{2mm}
			\If {$i > 0$}{
				$m$  = \clear{$m$, $i-1$}; // clear the previous bit \\
			}

			\vspace{2mm}
			$m$  = \set{$m$, $i$}; // set the i-th bit \\

			\vspace{2mm}
			\If{$m < a$}
			{
				 // try to make $m$ a little bigger \\
				$m$ = \set{$m$, $i+1$}\;
			}

			\vspace{2mm}
			$m_i$ = \refine{ $m, i+1, [a, b)$\,}\;

			\vspace{2mm}
			\If {$m_i \in [a,b)$} {
				$k = m_i$\;
			}
		}
		\textbf{return} $k$\;
	}
	\caption{pseudo-code for our algorithm}
\end{algorithm}

\RestyleAlgo{boxruled}
\begin{algorithm}%[frame=lines,label=inversion,caption={overall algorithm},captionpos=b]]
\SetStartEndCondition{ }{}{}%
\SetKwProg{Fn}{def}{\string:}{}
\SetKwFunction{Range}{range}%%
\SetKw{KwTo}{in}\SetKwFor{For}{for}{\string:}{}%
\SetKwIF{If}{ElseIf}{Else}{if}{:}{elif}{else:}{}%
\SetKwFor{While}{while}{:}{fintq}%
\AlgoDontDisplayBlockMarkers\SetAlgoNoEnd\SetAlgoNoLine%

	\SetKwFunction{ffs}{ffs}
	\SetKwFunction{clear}{clear}
	\SetKwFunction{clearPrefix}{clear\_prefix}
	\SetKwFunction{set}{set}
	\SetKwFunction{isSet}{is\_set}
	\SetKwInOut{Input}{Input}
	{
		\Input{ a range $[a,b)$, the number $m$ to refine, and the length $i$ of the prefix $m[0,i)$ to keep}

		\vspace{2mm}
		// jump to the next non-zero bit of $m$ \\
		i = \ffs{ \clearPrefix{$m, i$ }\,}\;

		\vspace{2mm}
		// check whether we are done \\
		\While{ $i < B-1$ }{

%			// jump to the next non-zero bit of $m$ \\
%			i = \ffs{ \clearPrefix{$m, i$ }\,}\;
%
%			\vspace{2mm}
%			// clear all the subsequent bits set \\
%			$k = m$\; 
%			\While{\isSet{m,i} }{
%				$k$ = \clear{$k, i$}\;
%				$i = i + 1$;
%			}
%
%			\vspace{2mm}
%			// set the first zero bit we found \\
%			k = \set{$k, i$}\;

			// jump to the next integer with a 0-bit at index i \\
			$k = m + (1 << i)$\;

			\vspace{2mm}
			// check whether it's a valid solution \\
			\If{ $k \in [a,b)$ }{
				// \refine{$k, i, [a,b)$ }\;
				m = k\;
			}

			\vspace{2mm}
			// jump to the next non-zero bit of $m$ \\
			i = \ffs{ \clearPrefix{$m, i + 1$ }\,}\;
		}
		\textbf{return} $m$\;
	}
	\caption{refine( $m, i, [a,b)$ )}
\end{algorithm}

\section{Alternative sampling schemes}

Another possibility to stratify across multiple dimensions would be to adopt the framework introduced by Gr\"{u}nschloss et al \shortcite{Gruenschloss:2012}.
This would be achieved considering a single global sequence $\phi_j$ and, for each leaf $i$, defining the $n$-th sample associated with that leaf by enumerating the $n$-th sequence sample falling into the interval $[i/N_l, (i+1)/N_l)$.
In this case, we would need to generalize the sample enumeration technique to finding the minimum across a range $[a,b)$ for a given sample index $n$.
In the following we will be assuming without lack of generality that that $N_l = 2^d$, as the other cases can be handled by taking a domain sized by the smallest power of 2 greater than $N_l$, and only focusing on the first $N_l$ entries.

Unfortunately, for this sampling scheme using a single 1D Halton sequence $\phi_j = \Phi_2$ would not be very useful.
In fact, the radical inverse is such that the $d$ least significant digits $l$ of an integer select an interval, while the higher bits specify the relative sample coordinates inside the interval.
In other words, with such a scheme a given leaf $i$ is addressed by $l$ = $\Phi_2^\leftarrow(i)$, while all the sequence points inside that leaf have index $l + n \cdot 2^d$, and the actual coordinates of the samples within the leaf depend on $n$ only, as we can see from the following equation:

\begin{equation}
\Phi_2(l + n \cdot 2^d) =  b^{-d} \cdot \Phi_2(n) + \Phi_2(l)
\end{equation}

This, however, means that for a given value of $n$, all the samples inside all leaves would be exactly the same: they would have relative coordinates $\Phi_2(n)$.
At that point returning the minimum over a given range $[a,b)$ would certainly be easy, as it would be a constant, but it wouldn't be of much use.

A more interesting alternative could be obtained by using a 2D Halton sequence, where the light vertex index $i$ is the first sequence dimension, and the eye vertex index $j$ is the second.
The sample enumeration framework would then allow us to ask for the sequence index of the $n$-th sample corresponding to any eye-light pair $(j,i)$.
In practice, the equation providing this index has the form:
\begin{equation}
\phi^\leftarrow(j,i,n) = (\Phi_2^\leftarrow(i) \cdot x + \Phi_3^\leftarrow(j) \cdot y \mod \, N_e N_l) + N_e N_l \cdot n
\end{equation}
where $x$ and $y$ are constants.
From the perspective of a fixed $j$, this can be further simplified to:
\begin{equation}
\phi_j^\leftarrow(i,n) = (\Phi_2^\leftarrow(i) \cdot x + c_j \mod \, N_e N_l) + N_e N_l \cdot n
\end{equation}

Finding the minimum value assumed by $\phi_j$ over a range $[a,b)$ would hence require the evaluation of:
\begin{equation}
\min_{i \in [a,b)} \left( \Phi_2( (\Phi_2^\leftarrow(i) \cdot x + c_j \mod \, N_e N_l) + N_e N_l \cdot n ) \right)
\end{equation}
Again, due to the bit reversal nature of $\Phi_2$, this means finding the index $i$ that maximizes the number of rightmost zeros in the argument $(\Phi_2^\leftarrow(i) \cdot x + c_j \mod \, N_e N_l)$, or more formally the index that makes the expression belong to the smallest possible set $I_{s_1...s_{B}}$, where the initial set $I$ is now defined as $I = \Phi_2^\leftarrow([a,b))$.
Unfortunately, this seems rather difficult due to the presence of the modulo arithmetic.
We believe, however, that the approach we suggested in the previous section might have more desirable blue-noise distribution and discrepancy properties.

%
%A better alternative would be the use of a single Sobol sequence:
%\begin{gather}
%\phi_j(i)
% =
%  \begin{pmatrix}
%   b^{-1} \\
%   b^{-2} \\
%  \cdot \\
%  \cdot \\
%  \cdot \\
%  \end{pmatrix}
%\left[
%	\bf{C}
%  \begin{pmatrix}
%   a_0(i) \\
%   a_1(i) \\
%  \cdot \\
%  \cdot \\
%  \cdot \\
%  \end{pmatrix}
%\right]
%\end{gather}
%where the matrix-vector multiplication happens over a finite field, e.g. $\mathbb{Z}_b$.

\bibliographystyle{acmsiggraph}
%\nocite{*}
\bibliography{main}

\begin{thebibliography}{\protect\citename{Gr\"{u}nschloss et~al\mbox{.} }2012}

\bibitem[\protect\citename{Chaitanya et~al\mbox{.} }2018]{Chaitanya:2018:MBPT}
{\sc Chaitanya, C. R.~A., Belcour, L., Hachisuka, T., Premoze, S., Pantaleoni,
  J., and Nowrouzezahrai, D.}
\newblock 2018.
\newblock {Matrix Bidirectional Path Tracing}.
\newblock In {\em Eurographics Symposium on Rendering - Experimental Ideas and
  Implementations}, The Eurographics Association, W.~Jakob and T.~Hachisuka,
  Eds.

\bibitem[\protect\citename{Georgiev and Fajardo }2016]{Georgiev:2016}
{\sc Georgiev, I., and Fajardo, M.}
\newblock 2016.
\newblock Blue-noise dithered sampling.
\newblock In {\em ACM SIGGRAPH 2016 Talks}, ACM, New York, NY, USA, SIGGRAPH
  '16, 35:1--35:1.

\bibitem[\protect\citename{Gr\"{u}nschloss et~al\mbox{.}
  }2012]{Gruenschloss:2012}
{\sc Gr\"{u}nschloss, L., Raab, M., and Keller, A.}
\newblock 2012.
\newblock {Enumerating Quasi-Monte Carlo Point Sequences in Elementary
  Intervals}.
\newblock {\em Springer Proceedings in Mathematics and Statistics 23\/} (01).

\bibitem[\protect\citename{Heitz and Belcour }2019]{Heitz:2019:MCBN}
{\sc Heitz, E., and Belcour, L.}
\newblock 2019.
\newblock {Distributing Monte Carlo Errors as a Blue Noise in Screen Space by
  Permuting Pixel Seeds Between Frames}.
\newblock In {\em Eurographics Symposium on Rendering}, The Eurographics
  Association.

\bibitem[\protect\citename{Niederreiter }1992]{Niederreiter:1992:RNG:130653}
{\sc Niederreiter, H.}
\newblock 1992.
\newblock {\em {Random Number Generation and quasi-Monte Carlo Methods}}.
\newblock Society for Industrial and Applied Mathematics, Philadelphia, PA,
  USA.

\bibitem[\protect\citename{Tokuyoshi and Harada }2019]{Tokuyoshi:2019:HRR}
{\sc Tokuyoshi, Y., and Harada, T.}
\newblock 2019.
\newblock {Hierarchical Russian Roulette for Vertex Connections}.
\newblock {\em ACM Trans. Graph. 38}, 4 (July), 36:1--36:12.

\bibitem[\protect\citename{Veach }1997]{Veach:PHD}
{\sc Veach, E.}
\newblock 1997.
\newblock {\em Robust Monte Carlo Methods for Light Transport Simulation}.
\newblock PhD thesis, Stanford University.

\end{thebibliography}

\end{document}